\documentclass[onecolumn,showpacs,preprintnumbers,amsmath,amssymb,nofootinbib]{revtex4}
\usepackage{graphicx}
\usepackage{epsfig}
\usepackage{dcolumn}
\usepackage{bm}
\usepackage[english]{babel}

\newcommand{\scN}{{\scriptscriptstyle N}}

\newcommand{\beq}{\begin{equation}}
\newcommand{\eeq}{\end{equation}}
\newcommand{\ga}{\gamma }
\newcommand{\si}{\sigma }
\newcommand{\ove}{\overline }

\begin{document}

\title{On the mass composition of primary cosmic rays in the energy
region $10^{15}-10^{16}$ eV}

\author{Yu.F. Novoseltsev$^1$}%
 \email{novoseltsev@inr.ru}
 \author{R.V. Novoseltseva$^1$}\
 \author{G.M. Vereshkov$^1$}
\email{gveresh@gmail.com}
 \affiliation{Institute for Nuclear Research of Russian Academy of Sciences}

\begin{abstract}

The method of a determination of the Primary Cosmic Ray mass
composition is presented. Data processing is based on the
theoretical model representing the integral muon multiplicity
spectrum as the superposition of the spectra corresponding to
different kinds of primary nuclei. The method consists of two
stages. At the first stage, the permissible intervals of primary
nuclei fractions $f_i$ are determined on the base of the EAS
spectrum vs the total number of muons ($E_\mu \ge $ 235 GeV). At the
second stage, the permissible intervals of $f_i$ are narrowed by
fitting procedure. We use the experimental data on high multiplicity
muon events ($n_\mu \ge$~114) collected at the Baksan underground
scintillation telescope. Within the framework of three components
(protons, helium and heavy nuclei), the mass composition in the
region $10^{15}-10^{16}$~eV has been defined: $f_p = 0.235\pm 0.02$,
$f_{He} = 0.290\pm 0.02$, $f_H = 0.475\pm 0.03$.

\end{abstract}


\maketitle

\section{Introduction}

Up to now the energy spectrum and the mass composition of primary
cosmic rays (CR) have been measured by direct methods (on satellites
or stratosphere balloons) up to energies of about $E_\scN\simeq 100$
TeV ($E_\scN$ is the primary nucleus energy). At higher energies the
information about CR is obtained with the help of indirect methods
which consist in a measurement of different parameters of extensive
air showers (EAS). Parameters relating to the energy spectrum are in
rather good agreement between each other $[1-11]$, while the data on
the mass composition are inconsistent enough (see table \ref{tab1}).
\begin{table}[ht]
 \caption{Some parameters of the CR energy spectrum and mass
 composition obtained in different experiments. The second
 column shows the publish year and reference.}
\label{tab1}
\begin{center}
\begin{tabular}{l c c c c l}
\hline
Detector  & Reference  & $E_k$,PeV & $\gamma _1$ & $\gamma _2$ &\hspace*{1mm} $<lnA>,(p+\alpha)$\\
\hline \hline MSU & 97,\cite{Atra, Fomi} & 3 & $2.7\pm $ & $3.1\pm
$ & \hspace*{1mm} $0.62\rightarrow 0.24\ (p+\alpha) $\\
\hline EAS - TOP & 98,\cite{Agli3} & 3 & $2.76\pm 0.03$ & $3.19\pm
0.06$ & \hspace*{1mm} $7.1\rightarrow 8.9\ (A_{eff})$\\
\hline HEGRA     & 97,\cite{Lind} & 2 & $2.60\pm 0.10$ & $3.00\pm
0.15$ & \hspace*{1mm} $0.60\rightarrow 0.45\ (p+\alpha)$\\
\hline HEGRA     & 98,\cite{Bern} &   & $2.60\pm 0.10$ & $3.00\pm
0.15$ &
\hspace*{1mm} $2.0\rightarrow 2.4\ (<ln A>)$\\
\hline HEGRA     & 99,\cite{Rohr} &3.4& $2.67\pm 0.03$ & $3.33\pm
0.40$ &
\hspace*{1mm} $0.55\rightarrow 0.48\ (p+\alpha)$\\
\hline KASCADE   & 99,\cite{Glasst} & 4 &  2.7           &  3.1 &
\hspace*{1mm} $0.70\rightarrow 0.50\ (p+\alpha)$\\
\hline CASA - MIA& 99,\cite{Glas2} & 3 & $2.66\pm 0.02$ & $3.00\pm
0.05$ &
\hspace*{1mm} $1.3\rightarrow 3\ \ (<ln A>)$\\
\hline DICE      & 00,\cite{Sword} & 3 & $2.66\pm 0.02$ & $3.00\pm
0.05$ &
\hspace*{1mm} $1.5\rightarrow 0.9\ (<ln A>)$\\
 & & & & & \hspace*{2mm}$0.60 - 0.50 - 0.65\ (p+\alpha)$\\
\hline CASA -    & 01,\cite{Fowl} &2-3& $2.72\pm 0.02$ & $2.95\pm
0.02$ &
\hspace*{1mm} 0.62 - 0.80 - 0.43 $(p+\alpha)$\\
 - BLANCA & & & & &\hspace*{1mm} 1.7 - 1.0 - 1.4\ $(<ln A>)$\\
\hline
\end{tabular}
\end{center}
\end{table}

To investigate the CR mass composition, the EAS parameters are
measured which have to be distinct for EAS produced by different
kinds of nuclei. These are the number of muons in EAS \ $n_\mu$, the
maximum depth in atmosphere $X_m$, fluctuations of the maximum depth
$\sigma (X_m)$, a steepness of lateral distribution of particles
near EAS core $\rho (r)$ etc.

The main problem of indirect methods of CR investigation is that the
information about both the energy spectrum and the mass composition
must be obtained from the same data sample. This leads hereto that
the determination of CR mass composition is very difficult problem.
There are numerous and various uncertainties related with an
interaction model and methods of measurement of EAS characteristics
(having, as a rule, considerable errors) and with the CR energy
spectrum. These difficulties lead to large spread of obtained
results. In Table~\ref{tab1}, the parameters of the CR energy
spectrum ($E_k$ is the "knee" energy, $\ga_1$ and $\ga_2$ are slope
exponents of the spectrum before and after the "knee") and the mass
composition ($<lnA>$ is the average logarithm of the number of
nucleons in a nucleus, $(p+\alpha)$ is the light nuclei fraction)
obtained in different experiments are presented. One can see that
the trend to weighting mass composition is violated by results
reported in \cite{Sword,Fowl}.

In the paper we present a method of the determination of CR mass
composition on the base of data on the EAS spectrum vs the total
number of high energy muons $I(n_\mu)$.

The paper is constructed as follows. In Section 2, entry conditions
are described. In Section 3, we present the method of the
determination of primary nuclei fractions intervals $\Delta f_i$,
which ensure an agreement with experimental data within the limits
of one standard deviation.  The realization of the method is shown
in Sections 4 and 5. In Section 6, intervals of $\Delta f_i$
determined at the first stage are narrowed by fitting procedure.
Sections 7 and 8 are Discussion and Conclusion.

\section{Initial conditions}
We shall use the data on high multiplicity muon events ($n_\mu \ge$
114) collected at the Baksan underground scintillation telescope
\cite{Chud}. In \cite{Voev} the muon multiplicity spectrum (i.e.,
the number $m$ of muons hitting the facility at unknown position of
EAS axis) at $m\ge $ 20 was measured at zenith angles $\theta \le
20^o$. The threshold energy of muons coming from this solid angle is
235 GeV.

It is known, the muon multiplicity spectrum depends on the facility
geometry and selection conditions of the experiment. This leads to
that multiplicity spectra cannot be compared with each other. It
would be better to present the data as a function of some invariant
variable which does not depend on experiment conditions. In our
opinion, the total number of muons in the EAS, $n_\mu$, can be
chosen as a such variable.

In papers \cite{Nov1,Nov2,Nov4} we developed the method of
recalculation from multiplicity spectrum to the EAS spectrum vs the
total number of muons, $I(n_\mu)$. The formulation of the task is
following: let F(m) be the integral multiplicity spectrum obtained
at a certain facility. Let us define the parameter $ \Delta (m)
={\overline {m_1}/{n_\mu}}$, which is the average fraction of muons
hitting the facility in the case when the latter is crossed by
$m_1\ge m$ muons. Assuming  then $n_\mu = m/{\Delta (m)}$, we will
obtain the integral spectrum of EAS vs the total number of muons
\begin{equation}
I(\ge n_\mu) = \frac{1}{G(m)}F(m),
\end{equation}
here $G(m)$ is the acceptance of the facility for a collection of
events with muon multiplicity $\ge m$. (It should be explained that
m has to be high enough, for example $m >$ 20.)

The numerical values of parameters $\Delta (m)$ and $G(m)$
calculated in \cite{Nov2,Nov4}  with regard to the real structure of
the facility are presented in Table \ref{tab2}. $N(\ge m)$ is the
experimental number of events with the muon multiplicity $\ge m$
\cite{Voev}. Nonmonotony of the $N(\ge m)$ is due to the different
exposure time, $T_{rec}$.
\begin{table}[ht]
 \caption{Numerical values of $\Delta (m)$ and $G(m)$.
The error in calculation of $\Delta (m)$ and $G(m)$ does not exceed
2-3$\%$. Non-integer values of m are obtained because of corrections
at trajectories reconstruction.}
\label{tab2}
\begin{center}
\begin{tabular}{|c|c|c|c|c|c|}
\hline \\
$ m $ & $N(\ge m)$  &\ \ $T_{rec}$, $10^6$ s\ \ &\ \
$\Delta (m) $\ \ & \ \ $G(m)$, $m^2\cdot sr$\ \ & $n_\mu$ \\
\hline 21.9 & 127 & 1.67 & 0.280 & 60.5 & 78.2 \\
\hline 32.9 & 547 & 19.33 & 0.289 & 57.9 & 113.9 \\
\hline 44.5 & 270 & 19.33 & 0.295 & 56.6 & 150.8 \\
\hline 56.5 & 164 & 19.33 & 0.299 & 54.8 & 188.6 \\
\hline 82.1 &  66 & 19.33 & 0.306 & 53.2 & 268.2 \\
\hline 124.9 & 49 & 41.93 & 0.313 & 51.6 & 399.3 \\
\hline 211.6 &  7 & 41.93 & 0.319 & 50.4 & 663.8 \\
\hline
\end{tabular}
\end{center}
  \vspace{-0.5pc}
\end{table}

The method developed is universal and allows to combine results
obtained in different experiments with muon bundles. In our case we
have combined the results reported in \cite{Voev} and
\cite{Nov4,Nov3} and obtained the EAS spectrum vs the total number
of muons in the range $75\le n_\mu \le 4000$, which corresponds to
the primary energy range of $10^{15}\le E_\scN \le 10^{17}$~eV
(Fig.1). It should be clarified that the data at $n_\mu > $ 2000 are
obtained for the muon threshold energy $E_{th} = $ 220 GeV, while
the points at $n_\mu < $ 700 have $E_{th} = $ 235 GeV which is the
threshold energy in the experiment \cite{Voev}. (We do not
recalculate the data to the same threshold energy to avoid
additional errors.)\ In Fig.1, the expected fluxes are calculated
for $E_{th} = $ 235 GeV ($n_\mu < $ 1000, dotted curves) and $E_{th}
= $ 220 GeV ($n_\mu
>$ 1000, solid and dashed curves). Numbers near curves denote the
mass composition variants: 1 is the low energy composition (the
nuclei fractions in percentage are 39, 24, 13, 13, 11), 2 is the
composition (\ref{I1}). Note there is no normalization in Fig.1.

The data at $n_\mu < $ 700 can be used for retrieval of information
on the CR mass composition in the region $E_\scN = 10^{15} -
10^{16}$~eV. Let us remark here that the data at $m = 124.9$ and $m
= 211.6$ in \cite{Voev} were obtained with essential systematic
errors: according to our estimates, the values of $m$ in these
points are underestimated 4\% and 10\% respectively \cite{Nov5},
therefore we restrict ourselves to the data at $m\le 82$ ($n_\mu < $
270).
\begin{figure}[t]
\epsfig{figure=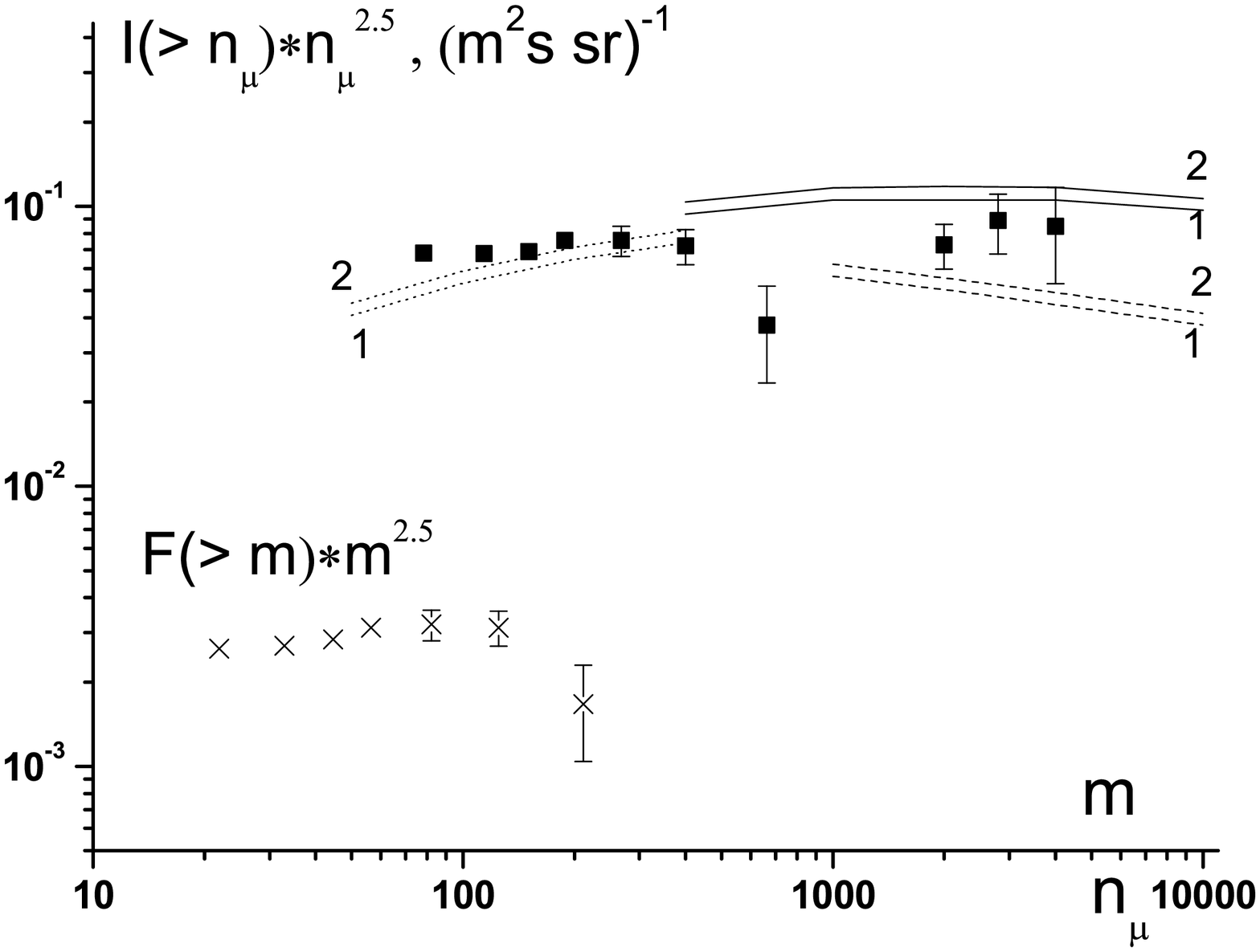,width=14.cm} \caption{Squares are the EAS
spectrum vs $n_\mu$ (experimental data). The muon threshold energy
is $E_{th} =$ 235~GeV if $n_\mu < $ 1000 and $E_{th} =$ 220 GeV at
$n_\mu > $ 1000 \cite{Nov1,Nov4}. Crosses show the muon multiplicity
spectrum obtained in \cite{Voev} ($m$ and $F(m)$ correspond to the
multiplicity spectrum). Solid curves are expected fluxes ($E_{th} =$
220 GeV) for the case $E_k = Z\cdot 3\cdot 10^{15}$ eV, dashed
curves -- the case $E_k = 3\cdot 10^{15}$ eV/nucleus. Dotted curves
show expected fluxes for the case $E_k = Z\cdot 3\cdot 10^{15}$ eV
at $E_{th} =$ 235~GeV. Numbers near curves denote the mass
composition variants: 1 is the "standard" (low energy) composition,
2~is the composition (\ref{I1}).} \label{fig_I1}
\end{figure}

As initial conditions, we use the mass composition obtained by
Swordy \cite{Sword1} with the help of compilation of results of
direct measurements at energies $\simeq $ 100~TeV per
nucleus\footnote{in comparison with the composition presented in
\cite{Sword1}, in (\ref{I1}) the proton fraction is increased by 5\%
(at the expense of helium nuclei) in accordance with data of
\cite{Asak}} (A is the number of nucleons in a nucleus)
\begin{equation}
\begin{array}{c}
$p$ \hspace{5mm} $He$\ \ \ $CNO$ \ \ \ $Ne-S$ \ \ \ $Fe$ \\
\hspace{-11mm} A\hspace{8mm} 1 \hspace{7mm} 4 \hspace{8mm} 14  \hspace{7mm} 28 \hspace{5mm} \ 56 \\
\hspace{-10mm} f,\% \ \  25 \hspace{5mm}  31 \hspace{7mm} 19 \hspace{6mm} 12 \hspace{7mm} 13 \\
\end{array}
\label{I1}
\end{equation}
and the proton flux at the energy  $E_p =$ 100 TeV measured in the
JACEE experiment \cite{Asak}.
\begin{equation}
 D_p(100\ TeV) = 2.95\times 10^{-10}\ (m^2\cdot s\cdot sr\cdot GeV)^{-1}
\label{I2}
\end{equation}
Then the total flux of nuclei with energy of $E =$ 100 TeV is equal
to $D_{tot} = D_p/0.25 = 11.8\times 10^{-10}\ (m^2\cdot s\cdot
sr\cdot GeV)^{-1}$, that is in a good agreement with the result
obtained at Tibet array \cite{Amen}. In this case, the mass
composition at the same energy per nucleon is
\begin{equation}
F(100\ TeV) = \left(
\begin{array}{c}
  0.88633 \\
  0.10412\\
  0.007584\\
  0.001474\\
  0.000492
\end{array}
\right) \label{I3}
\end{equation}
and the total flux of nuclei with the energy 100 TeV/nucleon is
equal to
\begin{equation}
 D_N(100\ TeV) = \sum_j D_{tot}f_jA^{-1.7}_j = 3.329\times 10^{-10}\ (m^2\cdot s\cdot sr\cdot
 GeV)^{-1}.
\label{I4}
\end{equation}
Our goal is to determine the mass composition evolution (from the
composition (\ref{I1})) into the region $E_\scN = 10^{15} -
10^{16}$~eV on the base of data on the multiplicity of high energy
muons ($E_\mu \ge $ 235 GeV) in EAS (see table \ref{tab2}). To this
end, we will use the measured fluxes of multiple muon events with
the multiplicity into differential intervals\ \ $n_{\mu i}\le n_\mu
\le n_{\mu (i+1)}$. At the first stage, we determine the permissible
intervals of primary nuclei fractions $f_i$ which ensure an
agreement with experimental data within the limits of one standard
deviation (Sections 4 and 5). And then, we refine the results with
the help of fitting procedure (Section 6).

To obtain the more certain results we fix the CR energy spectrum,
namely we adopt the conservative scenario:\\ i)~the slope change of
the spectrum occurs at the same energy per unit charge $E_k(Z) = $ 3
PeV$\times Z$,\\ ii) the spectra of all nuclei kinds have the slope
exponents $\ga_1 =$ 2.7 before the "knee"\ and $\ga_2 =$ 3.1 after
the "knee"
\begin{equation}
 D_A(E) = I_AE^{-2.7}(1 + E/E_k(Z))^{-0.4}.
\label{I5}
\end{equation}
As is seen from table \ref{tab1} this scenario is supported by
experimental data well enough.

It should be emphasized, we do not attempt to use the data at $n_\mu
>$ 2000 because the energy spectra of primary nuclei at $E_\scN
> 10^{16}$ eV are poorly understood.

\section{Equations}
The flux of events with muon multiplicity $n\ge n_\mu$ produced by
nuclei with A nucleons can be written in the form
\begin{equation}
\displaystyle I_A(n\ge n_\mu) = \int\limits_{E_{th}(A)}^\infty
D_A(E)P_A(E,\ n\ge n_\mu)dE\ ,
\label{01}
\end{equation}
here $D_A(E)$ is the differential flux of nuclei of kind A, $P_A(E,\
n\ge n_\mu)$ is the probability that the number of muons (with
$E_\mu \ge$ 235 GeV) in EAS produced by nucleus "A" (with energy $E$
per nucleon) is $n\ge n_\mu$, $E_{th}(A)$ is the threshold energy of
nuclei with A nucleons.

\begin{figure}[t]
\epsfig{figure=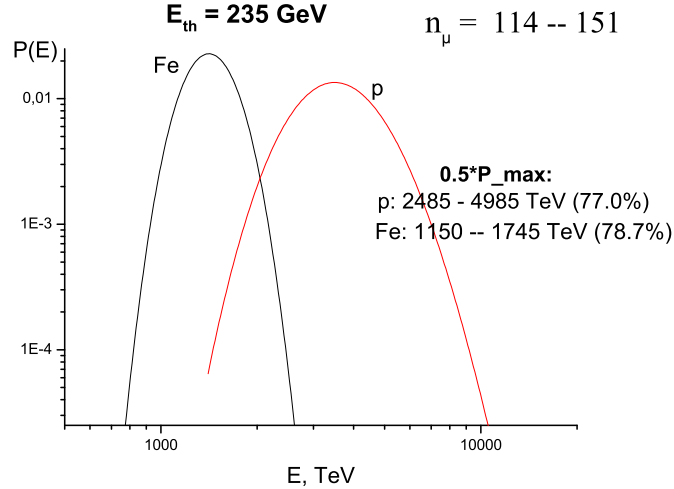,width=10.cm} \caption{Energy
distributions of protons and iron nuclei making a contribution to
the flux of muon events with $114\le n_\mu < 151$. Areas under
curves (p and Fe ) are equal to 1. The widths of distributions at
half-height and fractions of events in these regions are indicated.}
\label{fig1}
\end{figure}

We assume that the multiplicity of muons in EAS is described by the
negative binomial distribution $B_A(E,\ n)$ (Appendix B), then
\[
P_A(E,\ n\ge n_\mu) = \sum_{n\ge n_\mu} B_A(E,\ n)
\]
Taking into account that $D_A(E) = D_N(E)\times F_A(E)$, we rewrite
(\ref{01}) so
\begin{equation}
\displaystyle I_A(n\ge n_{\mu i}) = \int\limits_{E_{th}(A)}^\infty
F_A(E)D_N(E)P_A(E,\ n\ge n_{\mu i})dE, \label{002}
\end{equation}
and the flux of events with $n_{\mu i}\le n_\mu \le n_{\mu (i+1)}$
has the form
\begin{equation}
\begin{array}{c}
J_A(\Delta n_{\mu i}) = I_A(n\ge n_{\mu i}) - I_A(n\ge n_{\mu
(i+1)}) = \int\limits_{E_{th}^i(A)}^\infty
F_A(E)D_N(E)P_A(E,\ n\ge n_{\mu i})dE -\\
- \int\limits_{E_{th}^{i+1}(A)}^\infty F_A(E)D_N(E)P_A(E,\ n\ge
n_{\mu {i+1}})dE = \overline F_A R_{iA},
\end{array}
\label{02}
\end{equation}
where the first index of the matrix $R_{ij}$ points out to muon
multiplicity ($n_{\mu i}$) and the second one pertains to a nucleus
sort. $\overline F_A$ is the fraction of nuclei "A" averaged over
the energy region which gives the main contribution in the integral
(\ref{02}) (as is seen in Fig.\ref{fig1}, the region is rather
narrow). Thus we work in the approximation $F_j(E) \simeq \overline
F_j = const$ and will drop the symbol of averaging hereinafter.

A fast decrease of the CR flux with the energy is an important
simplifying factor. In consequence, the main contribution to the
muon bundles flux at any threshold multiplicity $n_\mu$ is
originated from nuclei whose energies are in a rather narrow region.
In Fig.\ref{fig1}, the energy distributions of protons and iron
nuclei making a contribution to the flux of events (EAS) with
$114\le n_\mu (E_\mu \ge 235\ GeV) < 151$ are shown. The widths of
distributions at half-height are $1745 - 1150$ TeV for iron nuclei
and $4985 - 2485$ TeV for protons.

To avoid possible methodical errors, we use only 4 points in the
spectrum $I_{tot}(\ge n_\mu) \equiv I(\ge n_\mu)$: at $n_\mu =$ 114,
151, 189, 268 (the point at $n_\mu =$ 78 has a different exposure
time and we do not use it in the present work (see table
\ref{tab2})). In table \ref{tab3}, the input data are presented:
muon multiplicity intervals $\Delta n_\mu $, the numbers and fluxes
of events in given intervals of~$n_\mu$.
\begin{table}[ht]
 \caption{Integral ($I$) and finite-difference ($J$) fluxes of events (EAS)
with the given number of muons ($E_\mu \ge$ 235 GeV). The flux
$J(\Delta n_\mu)$ is defined according to (\ref{02}).} \label{tab3}
\begin{center}
\begin{tabular}{|c|c|c|c|c|}
\hline \\
$n_\mu $ & $I(\ge n_\mu)\times 10^7$, $(m^2\cdot s\cdot sr)^{-1}$ & $ \Delta n_\mu $  & \ $N(\Delta n_\mu) $\ & \ $J\times 10^7$, $(m^2\cdot s\cdot sr)^{-1}$\ \\
\hline 114 & $4.887\pm 0.209$ & 114 - 151 & 277 & $2.419\pm 0.145$ \\
\hline 151 & $2.468\pm 0.150$ & 151 - 189 & 106 & $0.920\pm 0.089$ \\
\hline 189 & $1.548\pm 0.121$ & 189 - 268 & 98  & $0.906\pm 0.092$ \\
\hline 268 & $0.642\pm 0.079$ & $\ge$ 268 & 66  & $0.642\pm 0.079$ \\
\hline
\end{tabular}
\end{center}
  \vspace{-0.5pc}
\end{table}

We will solve a direct problem and define the regions of $F_j$
values which are compatible with equations (couplings)
\begin{equation}
 \begin{array}{c}
\displaystyle\ \ \sum_j R_{ij}\times F_j = J_i,\ \ (i=1,2,...,4)
\\[5mm]
 \end{array}
\label{2}
\end{equation}
where $J_i$ is the observed flux of events with muon multiplicity
from $i$-th interval -- $n_{\mu i}\le n_\mu < n_{\mu (i+1)}$.

Next we pass to the energy per nucleus and decrease the number of
independent variables with the help of relations
\begin{equation}
f_3(E) = f_4(E) = f_5(E),
\label{03}
\end{equation}
or
\begin{equation}
f_3(E) = 1.5f_4(E),\ \ f_4(E) = f_5(E).
\label{04}
\end{equation}
where $f_j(E)$ is the fraction of nuclei of kind j at the same
energy per a nucleus.

The relations (\ref{03}) are fulfilled at low energies ($E_\scN
\sim$ 100 GeV) and the relations (\ref{04}) are valid at $E_\scN
\simeq$ 100 TeV (see mass composition (\ref{I1})). We will find the
solution of equations (\ref{2}) in both cases, and in Section 7
discuss which variant ((\ref{03}) or (\ref{04})) is more preferable.

Thus we work in the approximation $f_3(E) \simeq f_4(E)\simeq
f_5(E)$ = 0.1467 and decrease the number of independent variables to
three: $f_1$, $f_2$, $f_3$.

Passing from five variables $F_j$ to three variables $f_k\
(k=1,2,3)$, it is convenient to rewrite the equations (\ref{2}) as
follows (we multiply and divide the j-th term by $A_j^{1.7}$ in each
equation):
\begin{equation}
\displaystyle \sum_{j=1}^3 R3_{ij}\times B_j = J_i,\ \ (i=1,2,...,4)
\label{4}
\end{equation}
where
\begin{equation}
\begin{array}{c}
R3_{ij} = R_{ij}/A_j^{1.7},\ \ \ B_j = F_j\times A_j^{1.7},\ j=1,2\\
R3_{i3} = \displaystyle \sum_{j=3}^5R_{ij}/A_j^{1.7},\ \ B_3 =
\displaystyle \frac{1}{3}\sum_{j=3}^5F_j\times A_j^{1.7}.
\end{array}
\label{4a}
\end{equation}
In addition
\begin{equation}
f_k = B_k\times \displaystyle \left[\sum_{j=1}^5F_j\times A_j^{1.7}
\right]^{-1}. \label{5}
\end{equation}

To determine $B_j$ we will use independent pairs of equations
(\ref{4}) (for example for i = 1,2 or i = 2,3 etc.), and for closure
of the equation system we use the normalization condition
$$f_1 + f_2 + 3f_3 = 1,$$
which (taking into account (\ref{4}), (\ref{5})) can be read so
\begin{equation}
B_1 + B_2 + 3B_3 = \sum_{j=1}^5F_j\times A_j^{1.7} \label{6}
\end{equation}

As it is known, an inverse problem is incorrect (in our case the
solution of system (\ref{4}) is unstable at small variations of
fluxes $J_i$). Therefore we solve the direct problem, namely: we define
the regions of variables $B_j$ which result in observed fluxes $J_i$
to within one standard deviation $\sigma_i = \sqrt J_i$. In the process
we use the mass composition (\ref{I1}) as initial conditions.

\section{Permissible domains (I)}
In this Section we illustrate the procedure of determination of
quantities  $B_j$ for the first two intervals of muon multiplicity:
$\Delta n_{\mu 1} = 114 - 151$ and $\Delta n_{\mu 2} = 151 - 189$
(see table~\ref{tab3}).

Let us write the equations (\ref{4}) for $i=1,2$ in an explicit form
\begin{equation}
\begin{array}{l}
\displaystyle \sum_{j=1}^3 R3_{1j}\times B_j = J_1,\\
\displaystyle \sum_{j=1}^3 R3_{2j}\times B_j = J_2,
\end{array} \label{08}
\end{equation}
or in a matrix form
\begin{equation}
R3\_1\times B = J\_1, \label{8}
\end{equation}
where $3\times 3$ matrix R3\_1  is composed of elements $R3_{ij}$
for i = 1,2, and the third row of the matrix is the normalization
conditions (\ref{6})
\begin{equation}
R3\_1 = \left(
\begin{array}{c c c}
  0.17374 & 0.38863 & 2.8885 \\
  0.073225 & 0.16861 & 1.3701 \\
   1 & 1 & 3
\end{array}
\right), \label{9}
\end{equation}
the index \_1 means that matrix $R3$ and vector $J$ correspond to
the first pair of equations (\ref{4}). Vector $J\_1$ is  by
definition (see table \ref{tab3})
\begin{equation}
J\_1 = \left(
\begin{array}{c}
  2.419 \\
  0.920 \\
  3.5453
\end{array}
\right) \label{10}
\end{equation}
(the third component of $J\_1$ is equal to the sum (\ref{6})
$\displaystyle J\_1_3 = \sum_{j=1}^5F_j\times A_j^{1.7}$).

With the help of (\ref{4}), (\ref{4a}) we get
\begin{equation}
F \rightarrow B^{in} = \left(
\begin{array}{c}
  0.88633 \\
  1.0991\\
  0.51997
\end{array}
\right) \label{11}
\end{equation}
where $B^{in}$ are the initial conditions.

Next we see that
\begin{equation}
R3\_1\times B^{in} = \left(
\begin{array}{c}
  2.0831 \\
  0.9626 \\
  3.5453
\end{array}
\right) \label{12}
\end{equation}

To determine the regions of $B_j$, which satisfy the relations
(\ref{08}) we shall make the mass composition heavier (or lighter)
$B^{in}\rightarrow B^{cur}$ until the result $R\times B^{cur} =
J^{cur}$ fall outside the limits $J_1 \pm \si_1$, or $J_2 \pm
\si_2$, where $\si_1 = \sqrt J_1$, $\si_2 = \sqrt J_2$ are the
errors of a flux measurement.

The change of the mass composition we shall realize by means of a
decrease (or an increase) of the proton fraction $B_1\rightarrow B_1
\mp \delta$ and an increase (or a decrease) of fractions of heavier
nuclei $B_2\rightarrow B_2 \pm \delta/4$, $B_3\rightarrow B_3 \pm
\delta/4$. Performing this procedure with a small step (for example
$\delta = 0.005$ ) we define the regions $$B_j^{low}\le B_j\le
B_j^{up},$$ which provide the
fulfillment of the conditions (\ref{08}) within one standard
deviation.

\begin{figure}[ht]
\epsfig{figure=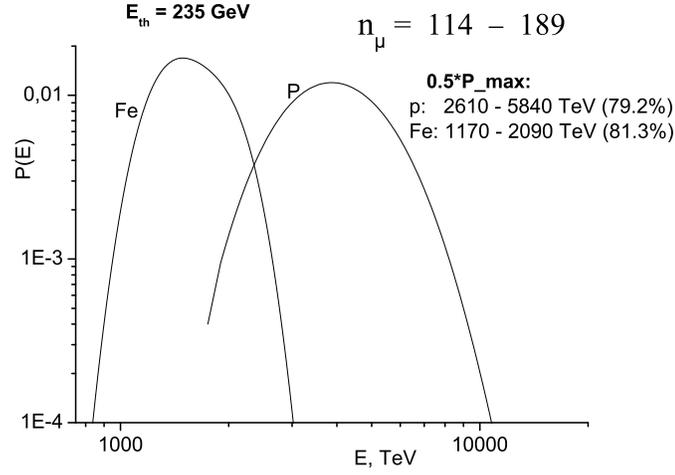,width=10.cm} \caption{Energy
distributions of protons and iron nuclei making a contribution to
the flux of muon events with $114\le n_\mu < 189$. Areas under
curves (p and Fe ) are equal to 1. The widths of distributions at
half-height and fractions of events in these regions are indicated.}
\label{fig2}
\end{figure}

The conditions
\begin{equation}
J_1 - \si_1 \le J^{cur}_1\le J_1 + \si_1,
\label{13}
\end{equation}
and
\begin{equation}
J_2 - \si_2 \le J^{cur}_2\le J_2 + \si_2.
\label{14}
\end{equation}
define different regions of $B_j$, of course. As we work in the
approximation of a slow change of the mass composition, one should
choose an intersection of the regions as the solution for the mass
composition averaged over the energy region under discussion ( $1170\le E_{Fe}\le 2090$
TeV, $2610\le E_p\le 5840$ TeV, see Fig. \ref{fig2}).

Note the choice of the common region for solution of equations
(\ref{08}) means the use of two experimental points simultaneously.
This reduces an influence of experimental errors.

It should be noted that the regions of $B_j$ values defined by
(\ref{13}), (\ref{14}) may be disjoint  at all. This would mean that
either i) the mass composition is changed very rapidly (and our
approximation $\overline F_j \simeq const$ is incorrect), or
ii)~experimental data have such large errors which do not allow the
simultaneous fulfillment of the conditions (\ref{13}), (\ref{14}).
The latter variant is more plausible and this should keep in mind in
what follows.
\begin{figure}[ht]
\includegraphics[width=10.cm]{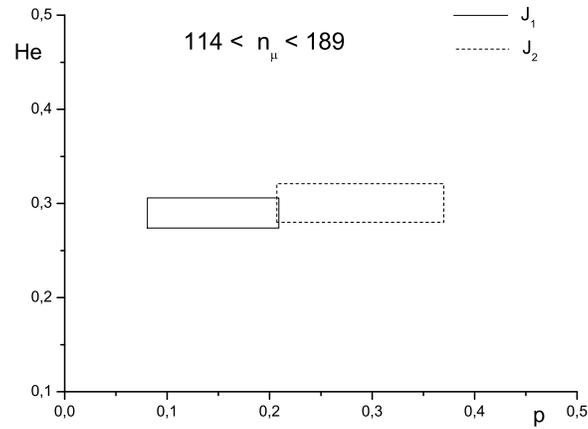}
\caption{Permissible domains for the protons and He nuclei fractions
according to the first pair of equations (\ref{4}), $i = 1,2$.}
\label{fig3}
\end{figure}

Solving equations (\ref{08}) ($i = 1,2$) for the conditions
(\ref{13}) we obtain (we present the results for variables $f_i$,
according to (\ref{5})):
\begin{equation}
\begin{array}{c}
  0.0808\le f_1\le 0.209 \\
  0.274\le f_2\le 0.306 \\
  0.172\le f_3\le 0.205
\end{array}
\label{15}
\end{equation}
and for conditions (\ref{14})
\begin{equation}
\begin{array}{c}
  0.207\le f_1\le 0.370 \\
  0.280\le f_2\le 0.321 \\
  0.117\le f_3\le 0.158
\end{array}
\label{16}
\end{equation}
The results are pictorially represented in Fig.\ref{fig3}. The intersection
of the regions (\ref{15}) è (\ref{16}) is:
\begin{equation}
\begin{array}{c}
  0.207\le f_1\le 0.209 \\
  0.280\le f_2\le 0.306 \\
  0.172\le f_3\le 0.158
\end{array}
\label{16a}
\end{equation}

As is obvious, the $f_3$ domains disjoint. We discuss a possible
reason of that in Section \ref{fit}. As the temporary solution, we
choose the average value -- $f_3$ = 0.165.

\begin{figure}[ht]
\includegraphics[width=10.cm]{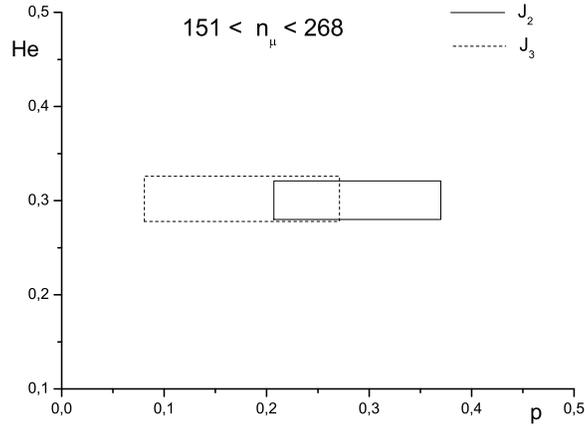}
\caption{Permissible domains for the protons and He nuclei fractions
according to the second pair of equations (\ref{4}), $i = 2,3$.}
\label{fig4}
\end{figure}

In a similar way, using the second pair of equations (\ref{4}) for $i =
2,3$ ($\Delta n_{\mu 2} = 151 - 189$ and $\Delta n_{\mu 3} = 189 - 268$), we get:\\

for the condition $J_2 - \si_2 \le J^{cur}_2\le J_2 + \si_2$
\begin{equation}
\begin{array}{c}
  0.207\le f_1\le 0.370 \\
  0.280\le f_2\le 0.321 \\
  0.117\le f_3\le 0.158
\end{array}
\label{17}
\end{equation}
that coincides with (\ref{16}), certainly, and for the condition
$J_3 - \si_3 \le
J^{cur}_3\le J_3 + \si_3$
\begin{equation}
\begin{array}{c}
  0.0808\le f_1\le 0.271 \\
  0.278\le f_2\le 0.326 \\
  0.150\le f_3\le 0.198
\end{array}
\label{18}
\end{equation}
The intersection of the regions (\ref{17}) and (\ref{18}) (Fig.\ref{fig4})
defines the mass composition  in the range $151 \le n_\mu \le 268$
($1630 \le E_{Fe} \le 3060$ TeV, $3870 \le E_p \le 8750$ TeV):
\begin{equation}
\begin{array}{c}
  0.207\le f_1\le 0.271 \\
  0.280\le f_2\le 0.321 \\
  0.150\le f_3\le 0.157.
\end{array}
\label{19}
\end{equation}

Finally, the third independent pair of equations ($i = 3,4$,
$\Delta n_{\mu 3} = 189 - 268$ and $\Delta n_{\mu 4}\ge 268 $) gives
the result (\ref{18}) for $J_3 - \si_3 \le J^{cur}_3\le J_3 + \si_3$,
and for $J_4 - \si_4 \le J^{cur}_4\le J_4 + \si_4$
\begin{equation}
\begin{array}{c}
  0.192\le f_1\le 0.376 \\
  0.279\le f_2\le 0.325 \\
  0.115\le f_3\le 0.161.
\end{array}
\label{20}
\end{equation}
The intersection of the regions (\ref{18}) and (\ref{20}) defines
permissible domains for $f_i$: \footnote{Note we used the integral
point $n_\mu \ge$ 268 as $\Delta n_{\mu 4}$. In this case, $R_{4j}$
is determined by expression (\ref{002}): $R_{4j} = I_j(n_\mu \ge
268)$. This has no affects on the final result.}
\begin{equation}
\begin{array}{c}
  0.192\le f_1\le 0.271 \\
  0.279\le f_2\le 0.325 \\
  0.150\le f_3\le 0.161.
\end{array}
\label{21}
\end{equation}
The widths of energy distributions (at half-height) of iron nuclei
and protons making a contribution to the flux of events with $189\le
n_\mu \le 380$ are $2140\le E_{Fe}\le 4270$ TeV and $5330\le E_p\le
12400$ TeV.

\section{Permissible domains (II)}

Now we repeat the calculations of the previous Section using
relationships $f_3(E) = 1.5\times f_4(E), \ f_4(E) = f_5(E)$ instead
of (\ref{03}).

Note that in this case
\begin{equation}
R3_{i3} = \displaystyle R_{i3}/A_3^{1.7} +
\frac{2}{3}(R_{i4}/A_4^{1.7} + R_{i5}/A_5^{1.7}),\ \ B_3 =
\displaystyle \frac{3}{7}\sum_{j=3}^5F_j\times A_j^{1.7}. \label{22}
\end{equation}
instead of (\ref{4a}) and the normalization condition reads as
follows (compare with (\ref{6}))
\begin{equation}
B_1 + B_2 + \frac{7}{3}B_3 = \sum_{j=1}^5F_j\times A_j^{1.7}
\label{23}
\end{equation}
(We assume $f_3(E) = 1.5\times f_4(E)$ i.e. $F_3 A_3^{1.7} = 1.5F_4
A_4^{1.7}$, therefore factors 3/7 and 7/3 appear in (\ref{22}) and
(\ref{23}))

The matrix R3\_1 and vector $B^{in}$ take the form:
\begin{equation}
R3\_1 = \left(
\begin{array}{c c c}
  0.17374 & 0.38863 & 2.1756 \\
  0.073225 & 0.16861 & 1.0268 \\
   1 & 1 & 2.3333
\end{array}
\right),\ \ B^{in} = \left(
\begin{array}{c}
  0.88633 \\
  1.0991\\
  0.66855
\end{array}
\right)\label{24}
\end{equation}
Performing the procedure described it the previous Section, we
obtain for $i= 1,2$\\
\hspace*{50mm}  $J_1 - \si_1 \le J^{cur}_1\le J_1 + \si_1$
\hspace*{5mm} $J_2 - \si_2 \le J^{cur}_2\le J_2 + \si_2$
\begin{equation}
\begin{array}{c}
  0.0554\le f_1\le 0.185 \\
  0.277\le f_2\le 0.307 \\
  0.230\le f_3\le 0.273
\end{array}
\hspace{10mm}
\begin{array}{c}
  0.182\le f_1\le 0.349 \\
  0.287\le f_2\le 0.326 \\
  0.156\le f_3\le 0.211.
\end{array}
\label{25}
\end{equation}
The intersection of the regions (\ref{25}) is
\begin{equation}
\begin{array}{c}
  0.182\le f_1\le 0.185 \\
  0.287\le f_2\le 0.307 \\
  0.230\le f_3\le 0.211.
\end{array}
\label{26}
\end{equation}
As with the approximation (\ref{03}), the $f_3$ domains disjoint. We
choose the average value  $f_3$ = 0.220 as the temporary solution.
Note within the framework of our approximation $f_4 = f_5 =
\frac{2}{3} f_3$ = 0.147.

Permissible domains for $i = 2,3$ è $i = 3,4$ are:\\
\hspace*{65mm}   $i = 2,3$ \hspace*{30mm} $i = 3,4$
\begin{equation}
\hspace{10mm}
\begin{array}{c}
  0.182\le f_1\le 0.250 \\
  0.287\le f_2\le 0.321 \\
  0.203\le f_3\le 0.211
\end{array}
\hspace{10mm}
\begin{array}{c}
  0.160\le f_1\le 0.250 \\
  0.287\le f_2\le 0.321 \\
  0.203\le f_3\le 0.218.
\end{array}
\label{27}
\end{equation}

\section{Fitting procedure \label{fit}}
Thus for the variant (\ref{04}) $f_3(E) = 1.5\times f_4(E),\ \
f_4(E) = f_5(E)$ (we shall call it "Model I" \ in the discussion
that follows), we have obtained the following permissible domains of $f_i$:\\
\hspace*{35mm} $i = 1,2$ \hspace*{30mm} $i = 2,3$ \hspace*{30mm} $i
= 3,4$
\begin{equation}
\begin{array}{c}
  0.182\le f_1\le 0.185 \\
  0.287\le f_2\le 0.307 \\
  f_3 = 0.22 \\
  f_4=f_5 = 0.147
\end{array}
\hspace{10mm}
\begin{array}{c}
  0.182\le f_1\le 0.250 \\
  0.287\le f_2\le 0.321 \\
  0.203\le f_3\le 0.211 \\
  0.135\le f_4=f_5\le 0.141
\end{array}
\hspace{10mm}
\begin{array}{c}
  0.160\le f_1\le 0.250 \\
  0.287\le f_2\le 0.321 \\
  0.203\le f_3\le 0.218 \\
  0.135\le f_4=f_5\le 0.145
\end{array}
\label{D1}
\end{equation}
and for the variant (\ref{03}) $f_3(E) = f_4(E) = f_5(E)$ ("Model II")\\
\hspace*{35mm} $i = 1,2$ \hspace*{30mm} $i = 2,3$ \hspace*{30mm} $i
= 3,4$
\begin{equation}
\begin{array}{c}
  0.207\le f_1\le 0.209 \\
  0.280\le f_2\le 0.306 \\
  f_3 = 0.165
\end{array}
\hspace{10mm}
\begin{array}{c}
  0.207\le f_1\le 0.271 \\
  0.280\le f_2\le 0.321 \\
  0.150\le f_3\le 0.157
\end{array}
\hspace{10mm}
\begin{array}{c}
  0.192\le f_1\le 0.271 \\
  0.279\le f_2\le 0.325 \\
  0.150\le f_3\le 0.161.
\end{array}
\label{D2}
\end{equation}
The obtained results have the status of permissible intervals of
fractions $f_i$.

The second stage of data processing consists in narrowing of
permissible intervals of~$f_i$. With this end in view, we carry out
the simultaneous fit of 4 integral points (see table \ref{tab3}) and
3 finite difference points:
\begin{equation}
 \begin{array}{c}
\displaystyle I_{n_\mu}\equiv I(\ge n_\mu)\ ,\qquad n_\mu=114,\quad
151,\quad 189,\quad 268\ ,
\\[5mm] \displaystyle
J_{n_\mu-n'_\mu}\equiv I(\ge n_\mu)-I(\ge n'_\mu)\ ,\qquad
n_\mu-n'_\mu=114-151,\quad 151-189,\quad 189-268\ .
 \end{array}
\label{D4}
\end{equation}

It will be now recalled that for $i= 1,2$ the $f_3$ domains do not
have common region in both variants. This may be associated with
experimental errors (see the remark before (\ref{15})) at a point
(or at some points). It is clear from the general reasoning that the
further from reality are experimental points, the smaller is the
domain overlap. The smoothing procedure is needed in such a
situation.

The analysis has shown that the value of $I(\ge 151) = 2.468$ is
incompatible with the assumption on a slow change of the mass
composition in the energy range under consideration. In the present
study we restrict the discussion to the correction of second
integral point ($I(\ge 151) = 2.468\pm 0.150$) and the more
exhaustive analysis will be presented in the later paper. Namely, we
increase the second point value by $\frac{1}{2}\sigma$:
\begin{equation}
I(\ge 151)\rightarrow 2.468+ 0.075 = 2.543.
\label{D3}
\end{equation}

This shift results in the common region of $f_3$ domains for $i=
1,2$ and  in so doing the permissible intervals of $f_i$ become very
close to the ones for $i= 2,3$ and $i= 3,4$. \footnote{The smoothing
procedure gives the same results.}

Next we perform fitting of 7 points (\ref{D4}) requiring that:\\
i) all data (7 points) are satisfied within the limits of
experimental errors ($\pm 1\sigma$),\\
ii) fitted parameters ($f_i$) are within permissible intervals.

Taking into account (\ref{D3}) the permissible intervals of $f_i$
(see (\ref{D1}), (\ref{D2})) take the form:
\begin{equation}
 \begin{array}{c}
 Model\ I:
\\[5mm]
\displaystyle 0.182\le f_p \le 0.250\ ,
\\[5mm] \displaystyle
0.287\le f_{He} \le 0.321\ ,
\\[5mm] \displaystyle
0.203\le f_N \le 0.211\ ,\qquad 0.135 \le f_{Si}=f_{Fe} \le 0.141\ .
 \end{array}
\label{D5}
\end{equation}
\begin{equation}
 \begin{array}{c}
 Model\ II:
\\[5mm]
\displaystyle 0.207\le f_p \le 0.271\ ,
\\[5mm] \displaystyle
0.280\le f_{He} \le 0.321\ ,
\\[5mm] \displaystyle
0.150\le f_N = f_{Si}=f_{Fe} \le 0.157\ .
 \end{array}
\label{D6}
\end{equation}
In the process of fitting, $f_{He}$ and $f_{N}$ have been chosen as
independent fitted parameters. The rest $f_i$ are calculated with
the help of expressions (\ref{04}) or (\ref{03}).

All possible sets of $f_i$ for the Model I are presented in Table
\ref{tab4}. For each admissible value of $f_{N}$ was found the
permissible interval of $f_{He}$. In Table \ref{tab4}, the mean and
boundary values of $f_{He}$ are shown for each admissible $f_{N}$.
Columns $I_j, J_i$ show the values of the fluxes (\ref{D4})
calculated at given values of $f_i$ ($I_j$ and $J_i$ have to be
between min and max values which are shown in the second row). All sets of
$f_i$ presented in Table are statistically equivalent and can be
written as follows:
\begin{equation}
 \begin{array}{c} \displaystyle
f_p=0.2365\pm 0.0025\ ,\ f_{He}=0.2895\pm 0.0025\ , \ f_N=0.2038\pm
0.0008\ , \\[3mm]
 f_{Si}=f_{Fe}=0.1356\pm 0.0007\ .
\end{array}
\label{D7}
\end{equation}

The fractions of light and heavy nuclei are
\begin{equation}
 \begin{array}{c} \displaystyle
f_{light}=f_p+f_{He}=0.526\pm 0.005\ , \qquad f_{heavy}=
f_N+f_{Si}+f_{Fe}=0.474\pm 0.003\ .
\end{array}
\label{D8}
\end{equation}

\begin{table}[ht]
 \caption{Possible sets of primary nuclei fractions $f_i$ for the Model I }
\label{tab4}\setlength{\extrarowheight}{2pt}
\begin{center}
\begin{tabular}{|c|c|c|c|c|c|c|c|c|c|c|c|c|c|}
 \hline
$\chi^2$&$f_p$&$f_{He}$&$f_N$&$f_{Si}$&$f_{Fe}$&$\langle\ln
A\rangle$&$I_{114}$&$I_{151}$&$I_{189}$&$I_{268}$&$J_{114-151}$&
$J_{151-189}$&$J_{189-268}$\\
 \hline
min & 0.182&0.287&0.203&0.1353&0.1353&$\quad$&4.678&2.393&1.427&0.563&2.199&0.906&0.815\\
max &0.250&0.321&0.211&0.1467&0.1467&$\quad$&5.096&2.693&1.669&0.721&2.489&1.084&0.998\\[5mm]
 \hline \hline
0.333&  0.239&  0.287&  0.203&  0.135&  0.135&  1.929&  4.788&  2.581&  1.562&  0.707&  2.208&  1.018&0.855\\
 \hline
0.313&  0.237&  0.290&  0.203&  0.135&  0.135&  1.933&  4.833&  2.605&  1.577&  0.714&  2.228&  1.028&0.863\\
 \hline
0.333&  0.234&  0.292&  0.203&  0.135&  0.135&  1.936&  4.878&  2.629&  1.592&  0.721&  2.249&  1.038&  0.871\\
 \hline
0.317&  0.237&  0.287&  0.204&  0.136&  0.136&  1.937&  4.845&  2.612&  1.581&  0.716&  2.233&  1.031&  0.865\\
 \hline
0.359&  0.234&  0.290&  0.204&  0.136&  0.136&  1.941&  4.900&  2.641&  1.599&  0.724&  2.259&  1.042&  0.875\\
 \hline
0.339&  0.236&  0.287&  0.2046& 0.1364& 0.1364& 1.941&  4.879&  2.630&  1.593&  0.721&  2.249&  1.038&  0.872\\
 \hline
\end{tabular}
\end{center}
\end{table}


In the same manner, we obtain for the Model II:
\begin{equation}
 \begin{array}{c} \displaystyle
f_p=0.2405\pm 0.0045\ , \ f_{He}=0.2985\pm 0.0145\ , \ f_N=f_{Si}=
f_{Fe}=0.1535\pm 0.0035\ .
\end{array}
\label{D9}
\end{equation}
\begin{equation}
 \begin{array}{c} \displaystyle
f_{light}=f_p+f_{He}=0.539\pm 0.019\ , \qquad
f_{heavy}=f_N+f_{Si}+f_{Fe}=0.461\pm 0.010\ .
\end{array}
\label{D10}
\end{equation}

\section{Discussion}
The procedure used at the first stage is based on the operation with
two equations in 2 variables.  To do this, it is necessary to set
(in addition to a normalization condition) two relations between
fractions $f_i$. We use the relations (\ref{03}) or (\ref{04}).
These additional relationships have a phenomenological nature of
course.

Note the results obtained at the first stage depend on the initial
conditions, but the second stage of data processing cancels this
dependence practically.

Let us remark also the method can be used under any energies (for
example $E_\scN > 10^{16}$ eV, $n_\mu >$ 1000) if the energy spectra
of primary nuclei will be known.

In regard to errors of $f_i$ determination, it should be noted the
following.

Experimental data (4 points of the integral spectrum and 3 points of
the finite-difference spectrum constructed from them) are well
described by a power function:
\begin{equation}
 \begin{array}{c} \displaystyle
I(n>n_\mu)=A\cdot (114/n_\mu)^m,\hspace{10mm} J(n_\mu) = A\cdot
(m/114)\cdot (114/n_\mu)^{(m+1)}
\\[3mm]
A = 4.9442 \pm 0.1357, \qquad m = 2.3695 \pm 0.0921.
\end{array}
\label{D11}
\end{equation}
As we can see, the relative error in spectrum parameters is about
4\%, while the relative error of the initial experimental data on
the integral spectrum is about 5--10\%. This decrease of the
relative error is due to matching data with the function and its
derivative to get the results (\ref{D11}) (A and m), while the
initial errors of 5--10\% are applicable only to the function.

Taking into account (\ref{D11}), we are sure that correct processing
of the experimental data must detect the mass composition with
relative errors in fractions $f_i$ under 5\%. Of course, these
errors can be bigger due to additional simplifying assumptions, e.g.
MODEL I or MODEL II. In any case, if errors over 5\% are indicated
in the final result, then the precision of the method used to
process the experimental data (i.e. the solution algorithm for the
ill-posed problem) is lower than the precision of these data.

We now discuss the results of the paper. Numbers after $\pm$ signs
in (\ref{D7}) and (\ref{D9}) are not the fitting errors. These
formulas are compact descriptions of restricted domains, whose full
description is given in Tables. As for the true errors of $f_i$, the
situation is as follows. The errors can be computed only after
determining ALL solutions, matched with the experimental data via
integral and finite difference spectra. The used algorithm is
mathematically very rigid, since it requires matching independent
solutions of three pairs of equations with the solution of seven
equations. This algorithm does not detect all possible solutions but
only a part of them. But the algorithm itself is compatible, so it
is possible to claim that the permissible intervals of $f_i$ are
expanded by at most 5\% of the determined ones. If we also take into
account ambiguity of the interaction model choice (appendix A), then
we get the error value about 10\%.

In this context MODEL I and MODEL II result in the same results:
\begin{equation}
f_p = 0.236\pm 0.020,\ f_{He} = 0.290\pm 0.020,\ f_H = 0.474\pm
0.030 \label{D12}
\end{equation}
In our opinion, MODEL I is more preferred for the following reasons.
First, the relations (\ref{04}) are satisfied for mass composition
(\ref{I1}) defined at $E = $ 100 TeV, i.e. in the neighboring energy
region. In the second place, the same relations are characteristic
for chemical composition of shells of II type supernova stars, which
(according to current concepts) are the sources of high energy
cosmic rays.

\begin{figure}[ht]
\includegraphics[width=11.cm]{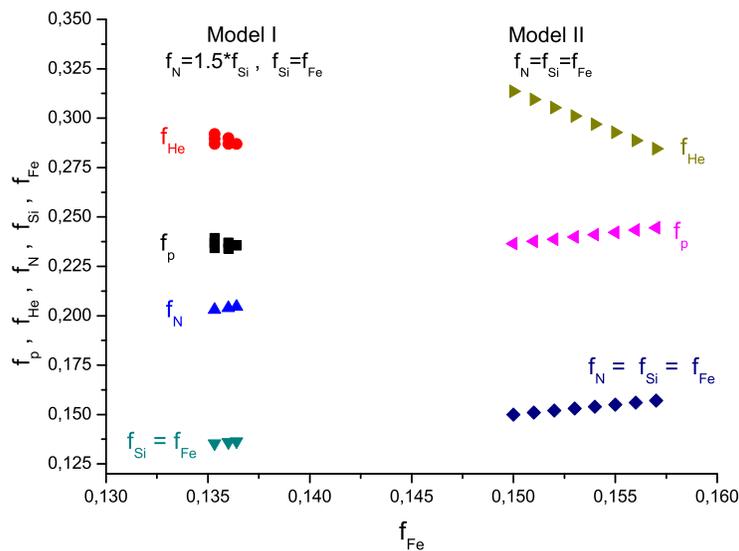}
\caption{Permissible domains for primary nuclei fractions $f_i$.}
\label{fig5}
\end{figure}

The results presented above are shown in Fig.\ref{fig5}. One can see
that uncertainty intervals of $f_i$ values for the Model I are
noticeably less than the ones for the Model II. Possibly this is one
more circumstance pointing to the most adequacy of the Model I.

\section{Conclusion}
We have presented the method of retrieval of information on CR mass
composition based on a solution of the direct problem (\ref{2}) with
a determination of permissible intervals of primary nuclei fractions
$f_i$. At the second stage, we constrict the intervals of $f_i$ with
the help of fitting procedure using the information about the
integral spectrum $I(\ge n_\mu)$ and its derivative $J(\Delta
n_\mu)$.

 Data processing is based on the theoretical model
representing the integral muon multiplicity spectrum as the
superposition of the spectra corresponding to different kinds of
primary nuclei. In so doing, it should be kept in mind that we have
fixed the interaction model (see appendix A).

With the framework of three components (protons, helium and heavy
nuclei) and under the assumption on a slowly change, the CR mass
composition in the region $10^{15}-10^{16}$~eV has been defined:
\begin{equation}
f_p = 0.236\pm 0.020,\ f_{He} = 0.290\pm 0.020,\ f_H = 0.474\pm
0.030 \label{D13}
\end{equation}
The result (\ref{D13}) should be read as the estimation (rather
precise) of CR mass composition in the energy region of
$10^{15}-10^{16}$~eV. Thus our analysis points out that CR mass
composition become some more heavy in comparison with the one
(\ref{I1}).

The method can be used under any energies if the energy spectra of
primary nuclei are known.

\section{Appendix A}
To calculate the parameters $\Delta (m)$ and $G(m)$
\cite{Nov1,Nov5}, we have used Monte-Carlo simulation and the
approximation of spatial-energy distribution function (SDF) of muons
in EAS obtained in \cite{Bozi},
$$\hspace*{30mm} f(r,\ge E,E_o ) = C\times exp{[-(r/r_o)^d]}, \hskip
45mm (A.1)$$
here $$\ \ r_o = \frac{0.95}{(1+12.5E)^{0.92}}+\frac{0.42}{E^{1.23}E_o^{0.9}}, \ \ \ d = 0.43 + \frac{0.2}{0.2+E_o},$$ $E_o$
is an energy per nucleon in a primary nucleus, $E$ is the muon
threshold energy (E in TeV, r in meters), $C$ - a normalization
factor.

The muon density $\rho$ at a distance r from EAS axis is difined by
the expression
$$\rho (r,\ge E,E_o)2\pi rdr = \overline{n}_\mu (A,E_o,\ge
E)f(r,E_o,\ge E)rdr, \hskip 45mm (A.2)$$
here $\overline{n}_\mu $ is the average number of muons with energy
$\ge E$ produced by a primary nucleus with energy $E_\scN = AE_o$ (A
is the number of nucleons in a nucleus) \cite{Bozi}:
$$\overline{n}_\mu (A,E_o,\ge E,\theta)=\frac{0.0187Y(\theta)A}{{E^a}}
\left(\frac{E_o}{E}\right)^{0.78}\left(\frac{E_o}{
{E_o+E}}\right)^{\delta} \hskip 45mm (A.3)$$ \noindent where $E_o$
and E in TeV,
$$a = 0.9+0.1lg(E),\ \ \delta = E+ \frac{11.3}{lg(10+0.5E_o)},
\ \ Y(\theta) = \frac {1+0.36\times ln(\cos\theta)}{\cos\theta}$$
$\theta $ - zenith angle.

\section{Appendix B}
The negative binomial distribution is a discrete probability
distribution of the number of successes in a sequence of Bernoulli
trials before a specified (non-random) number k of failures occurs
$$\hspace*{30mm} B(k,p) = C^{r+k-1}_r (1-p)^k p^r,\ \ k=0,1,2,... \hskip
45mm (B.1)$$
here $p$ is the probability of success.

Putting $r = n_\mu$ and taking into account $\ove n_\mu = kp/(1-p)$
we obtain
$$\hspace*{30mm} B(n_\mu , \ove n_\mu,k) = C^{n_\mu + k-1}_{n_\mu}
\left(\frac{\ove n_\mu /k}{\ove n_\mu /k + 1}\right)^{n_\mu}
\frac{1}{(\ove n_\mu /k + 1)^k} \hskip 25mm (B.2)$$ The parameter
$k$ was chosen in the form obtained in \cite{Bozi2}
$$\hspace*{30mm} k = \frac{\ove n_\mu}{f^2 - 1},\ \ \ f = A^{0.06}\left(1 + 0.013\frac{E_o}{E}\right)^{0.18}
\hskip 42mm (B.3)$$ At such choice of $k$ the variance of
$B(n_\mu,\overline n_\mu,k)$ distribution is greater than the one of
Poisson distribution in $6-10$ times for protons and $2-3$ times for
iron nuclei  in the energy region under discussion in accordance
with results of others works (e.g. \cite{Bilo,Atta}).

\section*{Acknowledgments}
This work was supported by RFBR grant 11-02-12043, the "Neutrino
Physics and Neutrino Astrophysics" Program for Basic Research of the
Presidium of the Russian Academy of Sciences and the Federal
Targeted Program of Ministry of Science and Education of Russian
Federation "Research and Development in Priority Fields for the
Development of Russia's Science and Technology Complex for
2007-2013", contract no.16.518.11.7072.

\end{document}